\documentclass{sig-alternate}

\usepackage{epsfig}
\usepackage{graphicx}
\usepackage{amsmath}
\usepackage{amssymb}
\usepackage{color}
\usepackage{graphicx}
\usepackage{booktabs}
\usepackage{fixltx2e}
\usepackage{subcaption}
\usepackage{tikz}
\usepackage{stmaryrd}
\usepackage[numbers]{natbib}
\usetikzlibrary{shapes.geometric, arrows}
\usepackage{xcolor}

\DeclareMathOperator* {\argmax}{arg\, max}

\makeatletter
\def\@copyrightspace{\relax}
\makeatother

\begin{document}

\title{Contextual Media Retrieval Using Natural Language Queries}

\author{Sreyasi Nag Chowdhury\hspace{1cm} Mateusz Malinowski\hspace{1cm} Andreas Bulling\hspace{1cm} Mario Fritz \\
\affaddr{Max Planck Institute for Informatics, Saarbr\"ucken, Germany}\\
\email{sreyasi,mmalinow,bulling,mfritz@mpi-inf.mpg.de}
}

\maketitle
\begin{abstract}
The widespread integration of cameras in hand-held and head-worn devices as well as the ability to share content online enables a large and diverse visual capture of the world that millions of users build up collectively every day. We envision these images as well as associated meta information, such as GPS coordinates and timestamps, to form a collective visual memory that can be queried while automatically taking the ever-changing context of mobile users into account. As a first step towards this vision, in this work we present \textit{Xplore-M-Ego}: a novel media retrieval system that allows users to query a dynamic database of images and videos using spatio-temporal natural language queries. We evaluate our system using a new dataset of real user queries as well as through a usability study. One key finding is that there is a considerable amount of inter-user variability, for example in the resolution of spatial relations in natural language utterances. We show that our retrieval system can cope with this variability using personalisation through an online learning-based retrieval formulation.
\end{abstract}






\section{Introduction}

Due to the widespread deployment of visual sensors in consumer products and Internet sharing platforms, we have collectively achieved a quite detailed visual capture of the world in space and time over the last years. In particular, mobile devices have changed the way we take pictures and new technology like life-logging devices will continue to do so in the future. With efficient search engines at our aid, viewing images and videos of unknown and distant places is just a few clicks away. These search engines do not allow for complex, natural language queries that include spatio-temporal references and they also largely ignore the users' local context.

\begin{figure}[t]
\centering
\begin{tabular}{@{}p{2.4cm}p{7cm}@{}}
\\
``What building is to the left of MPI-SWS?'' & \raisebox{-\totalheight}{\includegraphics[width=5.5em]{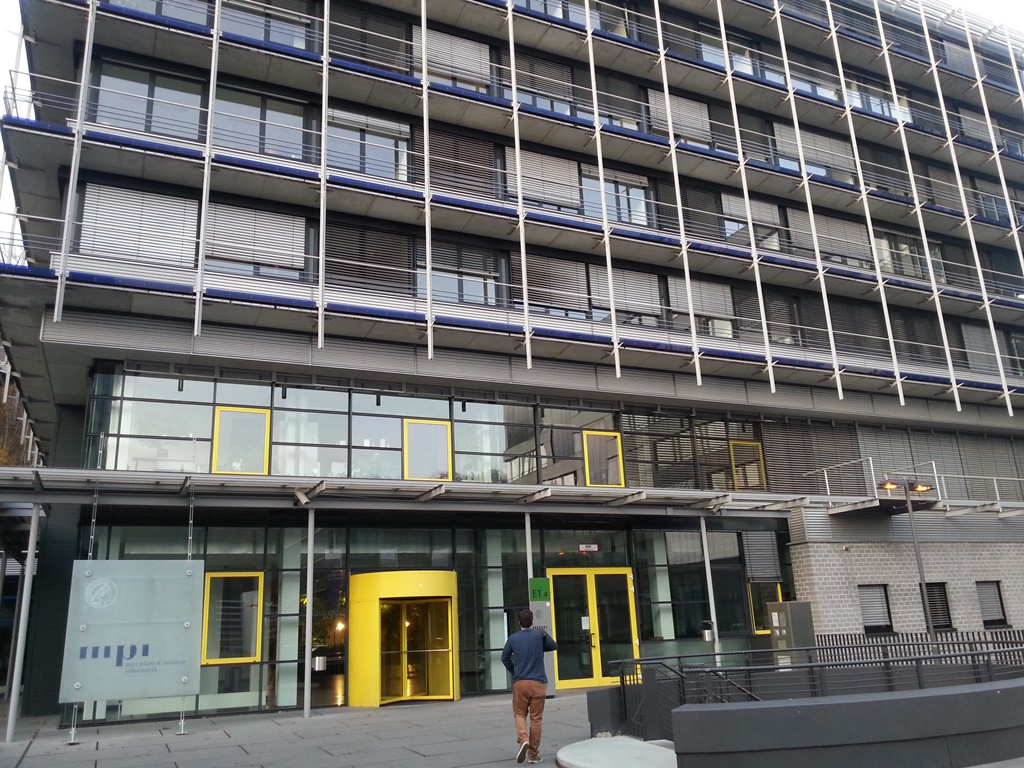}\hspace{0.5em}\includegraphics[width=5.5em]{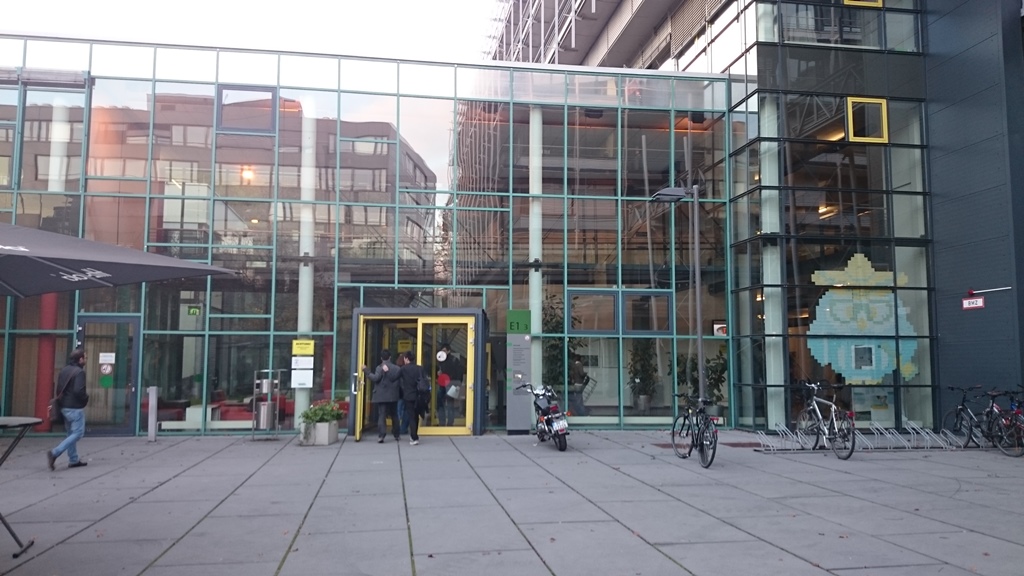}\hspace{0.5em}\includegraphics[width=5.5em]{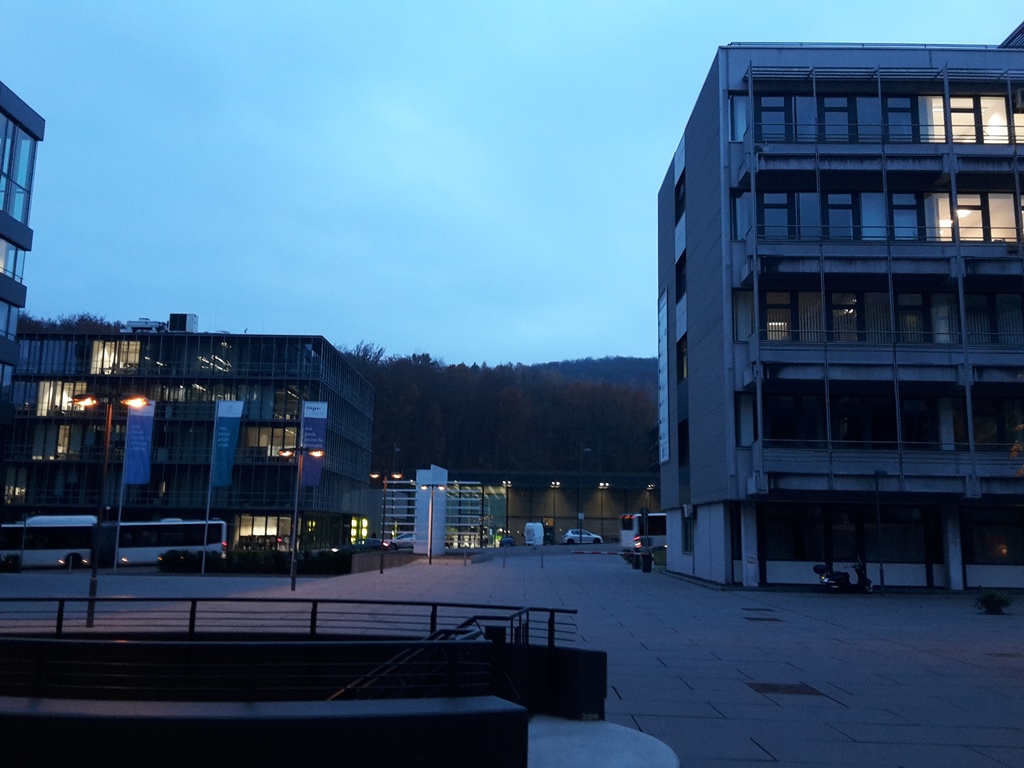}} \\
``What is in front of bus terminal?'' & 
\raisebox{-\totalheight}{\includegraphics[width=5.5em]{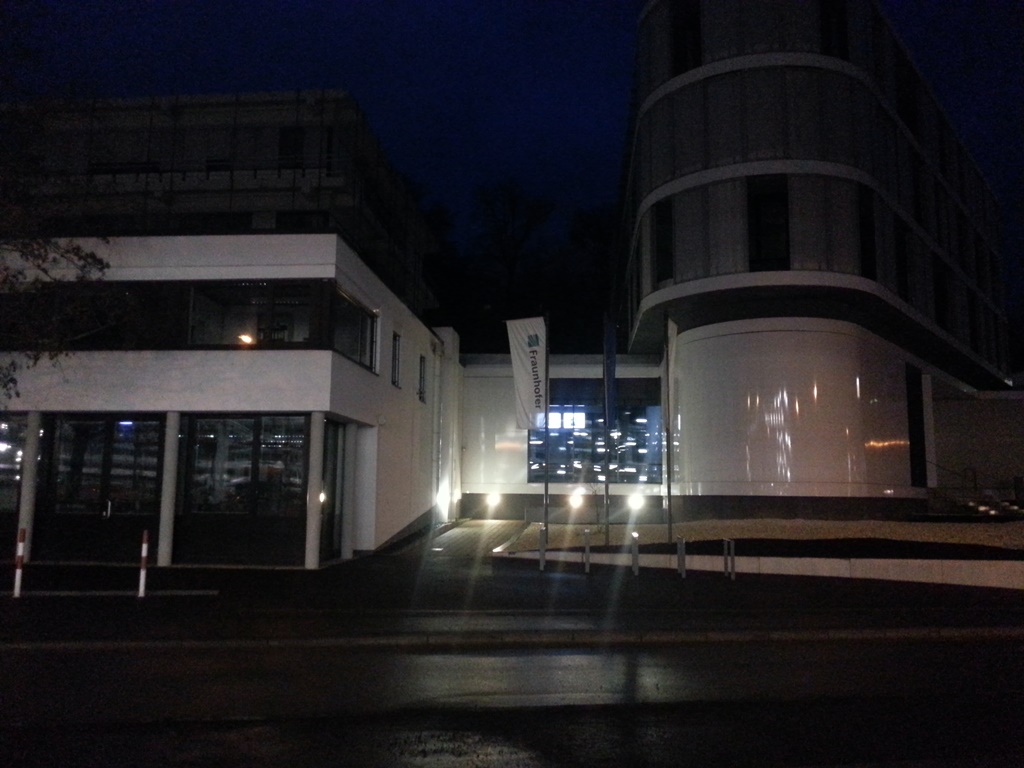}\hspace{0.5em}\includegraphics[width=5.5em]{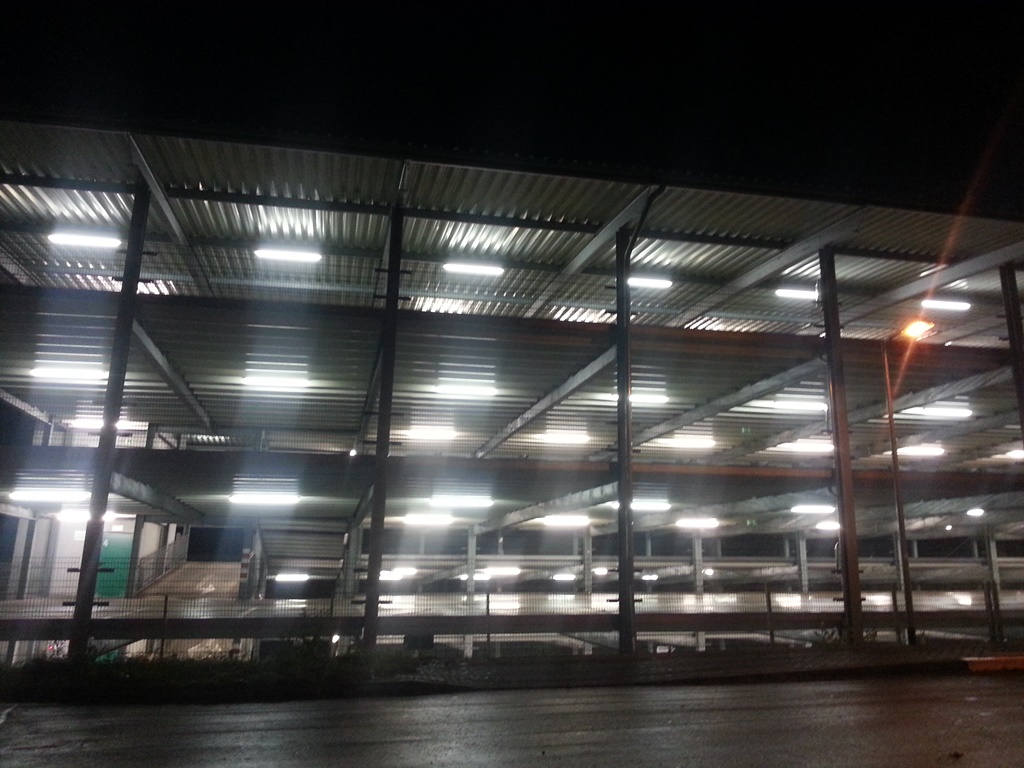}\hspace{0.5em}\includegraphics[width=5.5em]{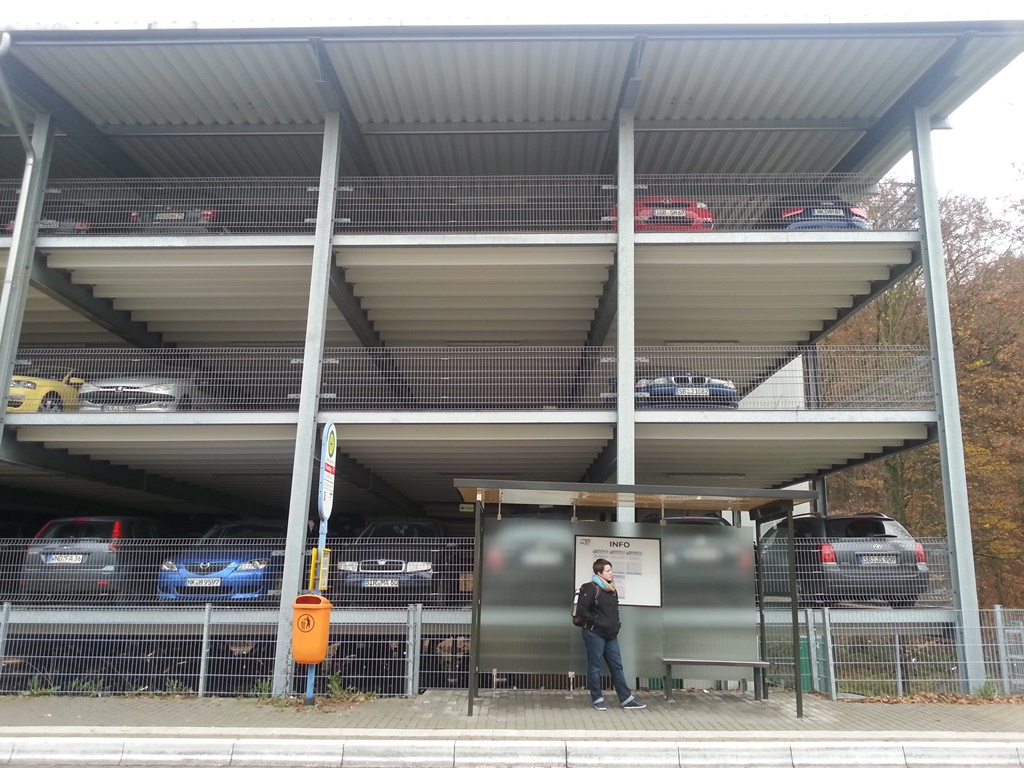}} \\
``What is near MPI-INF?'' & 
\raisebox{-\totalheight}{\includegraphics[width=5.5em]{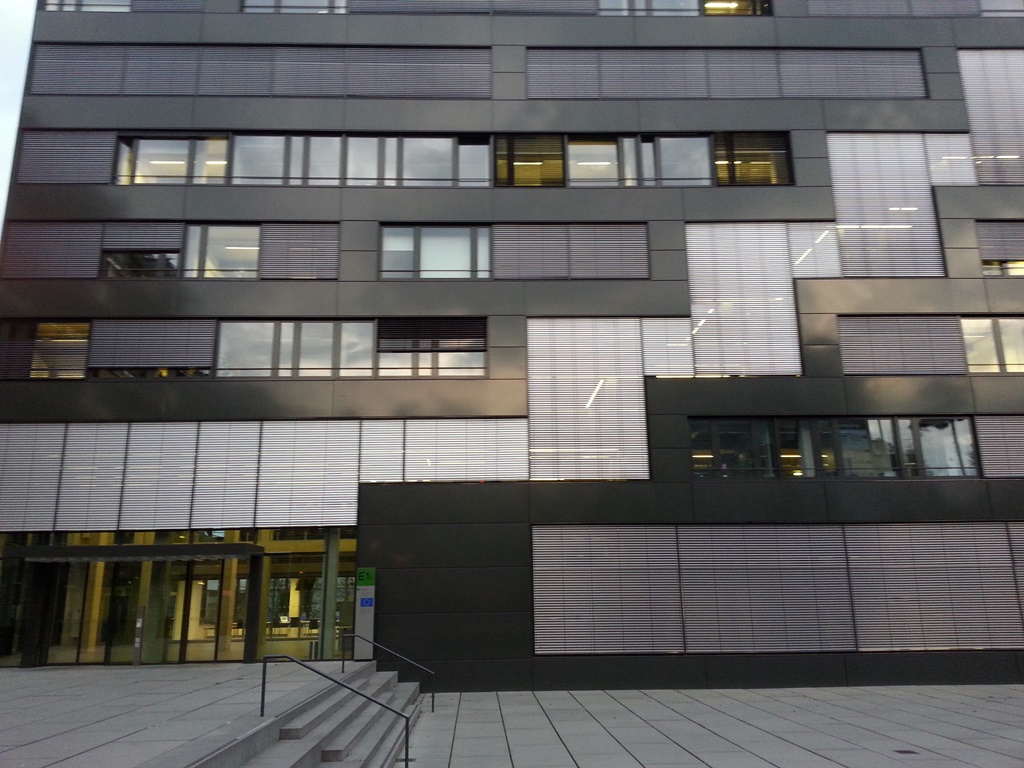}\hspace{0.5em}\includegraphics[width=5.5em]{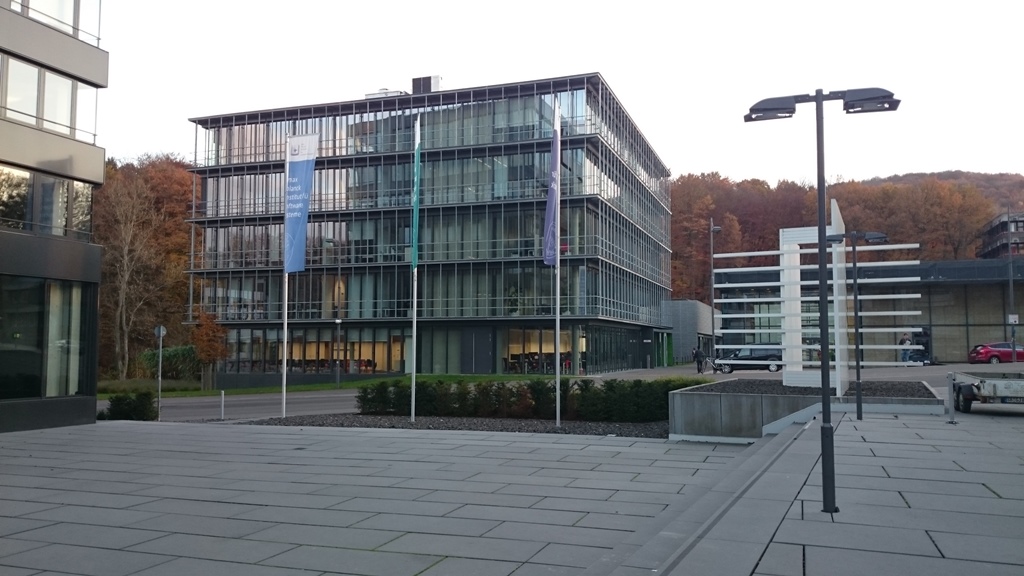}\hspace{0.5em}\includegraphics[width=5.5em]{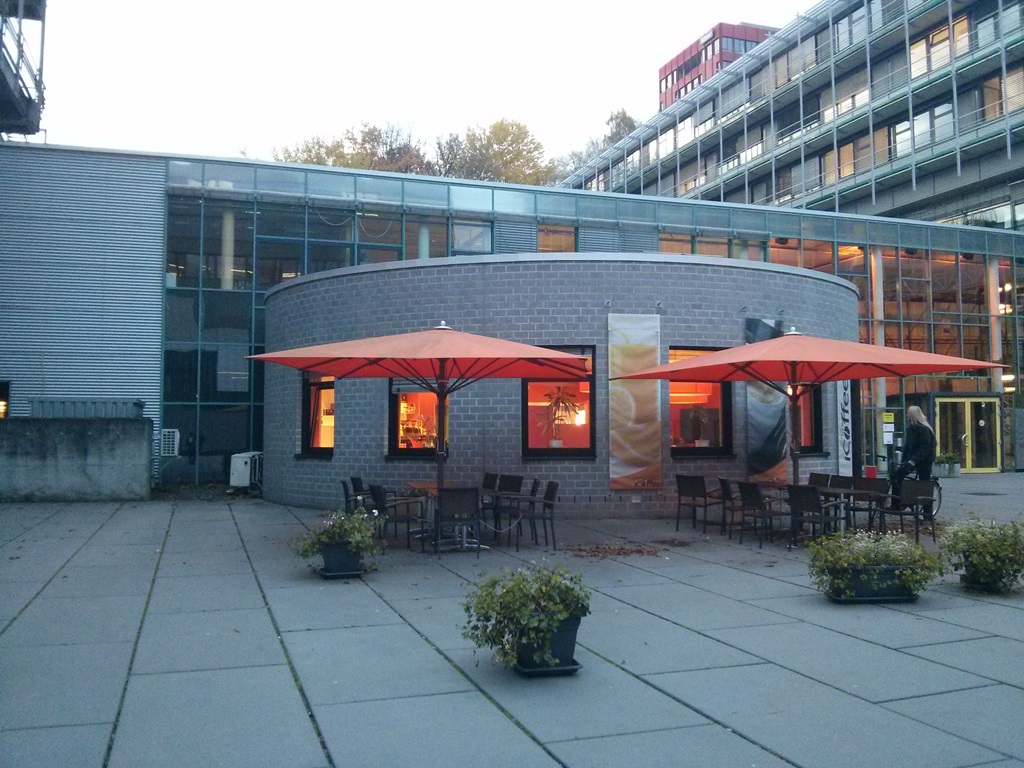}} \\
``What did this place look like in December?'' & 
\raisebox{-\totalheight}{\includegraphics[width=5.5em]{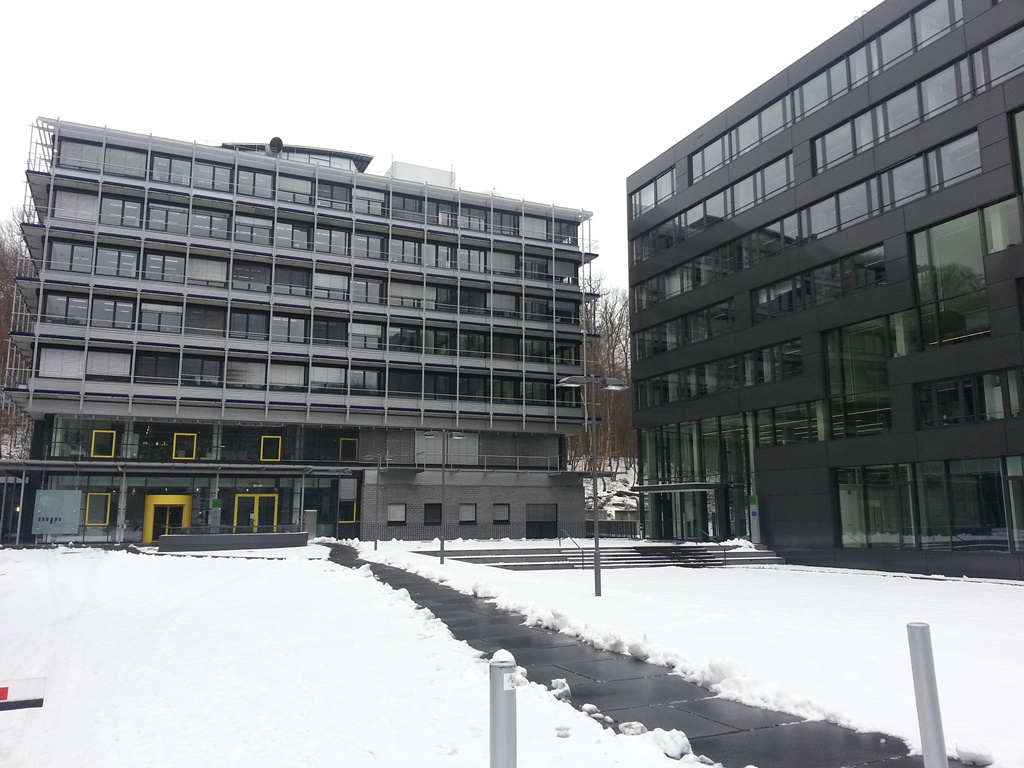}\hspace{0.5em}\includegraphics[width=5.5em]{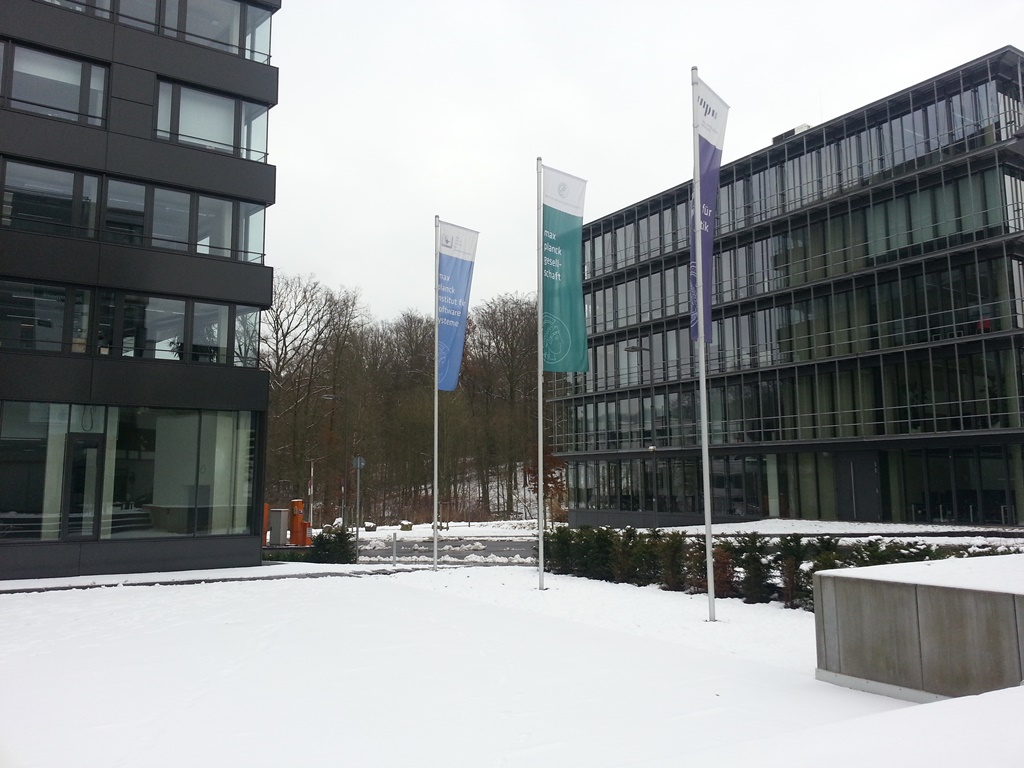}\hspace{0.5em}\includegraphics[width=5.5em]{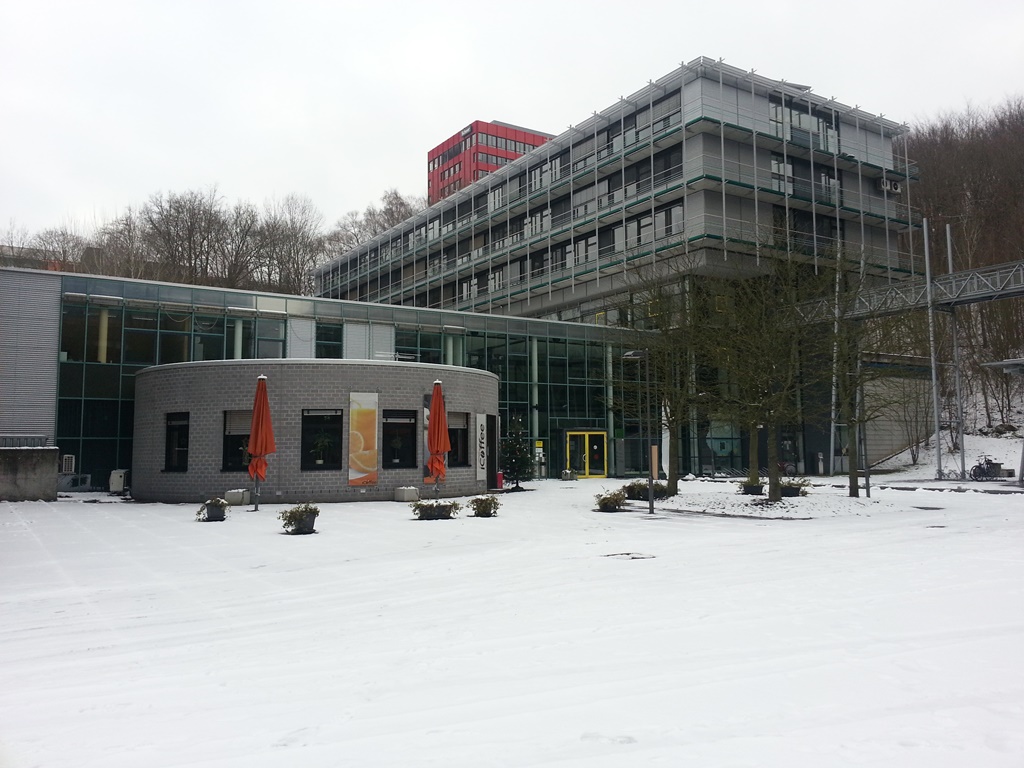}} \\
\end{tabular}
\caption{Sample queries and retrieved images of our contextual media retrieval system Xplore-M-Ego.}
\label{qualResults}
\end{figure}

Similar to how mobile devices have changed the way we take pictures, we ask how media search should be transformed to make use of the rich context available at query time. What if we quickly want to know what is behind the building in front of us? What if we want to know what a particular cafe looks like to quickly locate it in a busy market area? What if we want to see what our new neighborhood looks like in winter? Our approach makes use of the user's ever-changing context to retrieve results of a spatio-temporal query on a mobile device. The user is enabled to make references to his/her changing environment by allowing for queries in natural language. We have named our system Xplore-M-Ego (read \textit{``Explore Amigo''}) -- which stands for Exploration(Xplore) of Media(M) Egocentrically(Ego)".

\section{Related Work}

Previous research addressing the problems of media retrieval and machine understanding of natural language queries can be broadly classified into three groups.\\

\noindent
{\bf Spatio-temporal Media Retrieval:} 
Spatio-temporal media retrieval is the browsing of media content captured in different geographical locations at various times in the past. \citet{Snavely2006} proposed \textit{Photo tourism} that constructs a sparse 3D geometric representation of the underlying scene from images. Using this system users could move in 3D space from one picture to another. 
To address challenges in the construction management industry, \citet{Wu2009} designed \textit{PhotoScope}. It is an interactive tool to visualize spatio-temporal coverage of photos in a photo collection, which users can browse with space, time and standardized content specifications. 
\citet{tompkin2013video} developed \textit{Vidicontexts} that embeds videos in a panoramic frame of reference (context) and enables simultaneous visualization of videos in different foci. A similar system, \textit{VideoScapes} \cite{Tompkin2012} was implemented as a graph with videos as edges and portals (automatically identified transition opportunities) as vertices. When temporal context was relevant, videos were temporally aligned to offer correctly ordered transitions. 

In contrast to these methods, our approach implements egocentrism by taking users' context (geographical location and viewing direction) into account, and interfaces with the users through natural language queries.\\


\noindent
{\bf Natural Language Query Processing:}
Successful answering of a natural language question by machines requires understanding its meaning, which is often realizable by a semantic parser that transforms the question into its formal representation. Traditional approaches to semantic parsing used supervised learning  by training on questions with costly manually annotated logical forms~\cite{zettlemoyer2012learning, wong2007learning}. Modern approaches use more scalable techniques to train a semantic parser with more accessible textual question-answer pairs~\cite{clarke2010driving,Liang2012,berant2014semantic}. \citet{Malinowski2014} proposed an architecture for question-answering based on real-world indoor images. They  extended the work of \citet{Liang2012} to include subjective interpretations of scenes. They also identified challenges that holistic architectures have to face, such as different frame of reference in spatial relations or ambiguities in the answers~\cite{Malinowski2014,Malinowski2014a}.

Our work differs in that we target a dynamic and egocentric environment in contrast to static geographical/job/image data.\\


\noindent
{\bf Media Retrieval Using Natural Language Queries:}
Previous research on media retrieval using natural language queries varies considerably
in the methods used to process the natural language utterances.  \citet{lum1992intelligent} presented a method that matches semantic network representations of queries with those of natural language descriptions of media data (manually annotated). \citet{kucuktunc2007natural} proposed a pattern matching approach based on Part-of-Speech (POS) tags. Other approaches are based on RDF-triples \cite{hwang2007method} and SPARQL queries~\cite{Hakeem2009}. Contrary to these research threads, our work does not involve any human annotations or additional processing steps for extracting descriptions of entities from images and videos. Instead, we extract media content simply based on its meta data such as geographical location (GPS coordinates) and textual questions.

Prior research also looked into media retrieval with natural language questions containing spatial relations. \citet{tellex2009towards} explored spatial relations in surveillance videos by a classification task which handles two prepositions, ``across'' and ``along''. \citet{lan2012image} used structured queries that consists of two objects linked by a spatial relations chosen from a restricted set of spatial prepositions. In contrast, our media retrieval architecture aims to operate on rich natural language questions that liberate from any artificially imposed restrictions, such as fixed structure of the questions or a restricted vocabulary. 

To the best of our knowledge, none of the previous works ventured into contextual media retrieval by taking into account the user's current location and viewing direction. The introduction of egocentrism and natural language queries in architectures developed for browsing large media collections have many practical applications. Not only does it open another unexplored dimension for media retrieval (vis-\`a-vis, ``egocentrism''), but also aids in human interaction with the computer.

\section{Contextual Media Retrieval}

Our contextual media retrieval architecture allows users to explore a collective media collections in a spatio-temporal context through natural language questions such as ``What is there in front of the university bus terminal?'', ``What is there to the left of the campus center?'', ``What happened here five days ago?'', ``What did this place look like in December?'' etc. In the following, we show how we formulate our architecture.
Particular attention is payed on how to cope with the user's dynamic context and spatial references in natural language questions. We further describe how we collect a data set which relate to our initial motivation of building a \textit{Collective Visual Memory}.

\hspace*{-1em}
\begin{figure*}
\begin{tikzpicture} [node distance=1.5em]
	\draw[thick] (0,-1) -- (6.75,-1) node[pos=0.5,above] {\scriptsize{natural language question}};
	\draw[thick,->] (0,-1) -- (0,-3.80);  
	\draw[thick,fill=gray!50] (6.75,-1.2) rectangle (7.75,-0.8) node[pos=0.5] {$user$};
	\draw[thick] (7.25,-1.2) -- (7.25,-2.8) node[pos=0.5, left] {\scriptsize{\rotatebox{90}{metadata}}};
	\draw[thick] (7.25,-2.8) -- (11.6,-2.8);
	\draw[thick,->] (11.6,-2.8) -- (11.82,-3.04);
	\draw[text width=3.9cm] (7.4,-2.09) rectangle (11.5,-2.7) node[pos=0.5] {\scriptsize{\texttt{\hspace{0.7em}user(GPS\_lat,GPS\_lon,view\_dir) \\ }}};
	
    \draw[white, text width=4cm] (17.2,-1.2) rectangle (18,-0.8) node[pos=0.5] {\includegraphics[width=1.2cm]{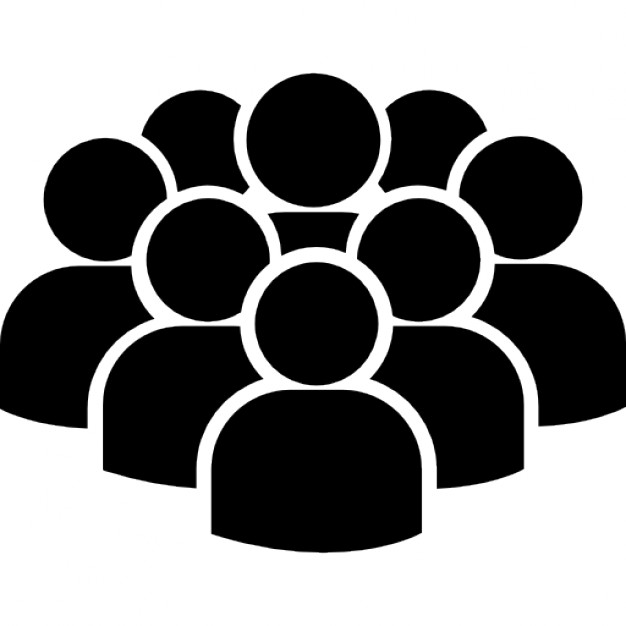}};
    \draw[thick, <-] (14, -1) -- (15.5, -1) node[pos=0.5,above] {\scriptsize{media}};
	\draw[white, text width=4cm] (10.8,-1) rectangle (14,-1) node[pos=0.5] {\includegraphics[width=1cm] {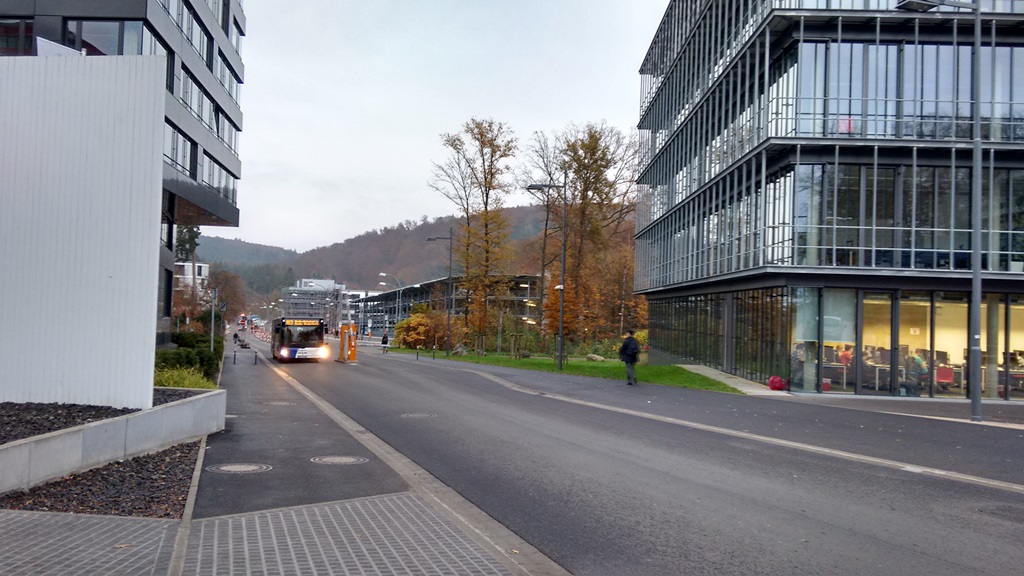}\hspace{0.15cm}\includegraphics[width=1cm] {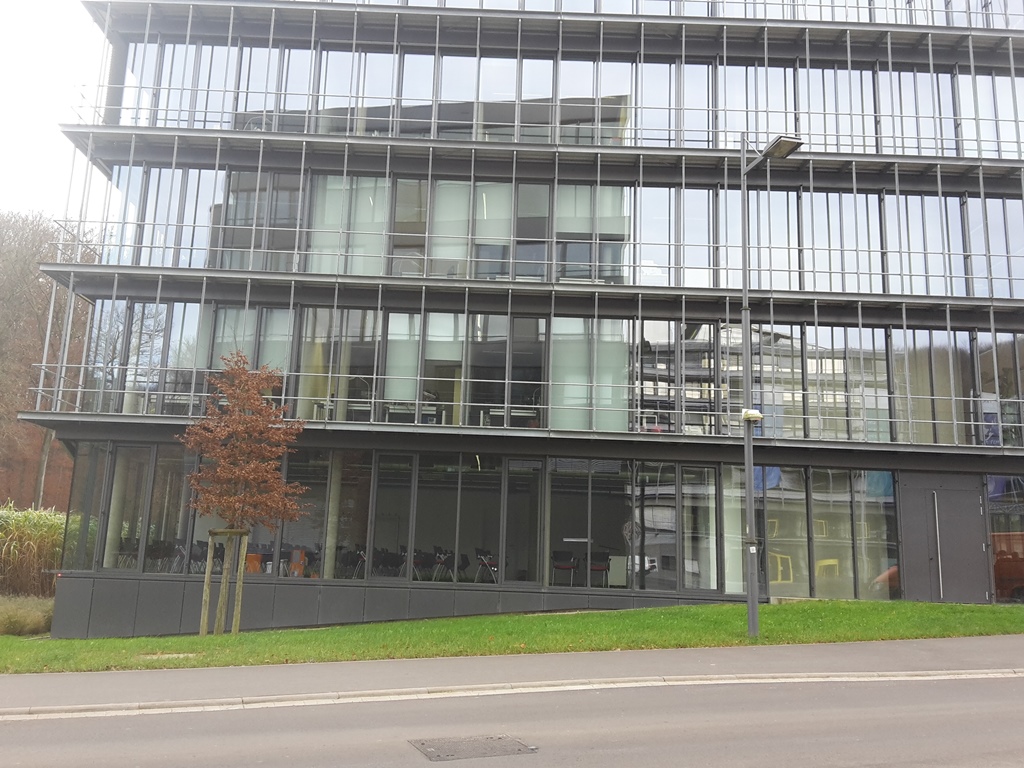}\hspace{0.15cm}\includegraphics[width=1cm] {figures/leftOfMpiSws2}\\\vspace{0.15cm} \includegraphics[width=1cm] {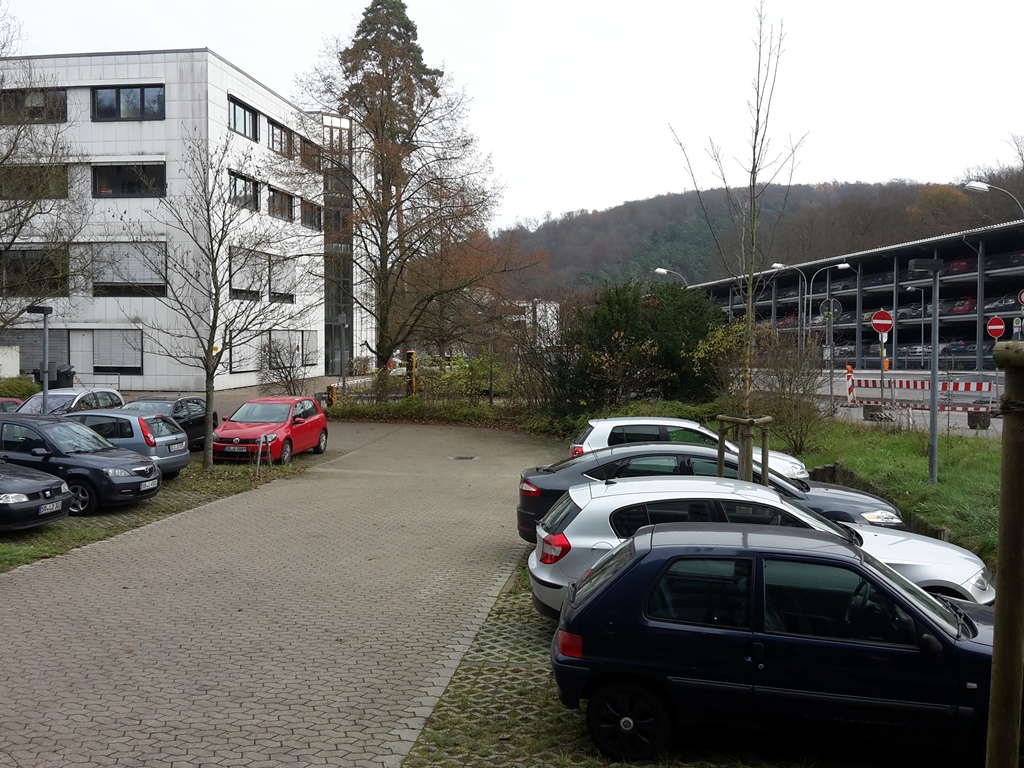}\hspace{0.15cm}\includegraphics[width=1cm] {figures/nearMpiInf3}\hspace{0.15cm}\includegraphics[width=1cm]{figures/frontOfBusTerminal3}};
	\draw[white, text width=1.5cm] (9.4,-1) rectangle (10,-1) node[pos=0.5] {\scriptsize{\textit{\color{black}Collective Visual Memory}}};
	\draw[double,->] (12,-1.95) -- (12,-2.95);
	\draw[text width=4.5cm] (12.2,-2) rectangle (16.7,-2.8) node[pos=0.5] {\scriptsize{\texttt{\hspace{0.6em}image(name,GPS\_lat,GPS\_lon,month)\\\hspace{0.6em}video(name,GPS\_lat,GPS\_lon,month)}}};
	
	\draw[thick,fill=gray!50] (12,-3.2) circle (0.25cm) node (wd) {$w_d$};
	\draw[thick]  (10,-3.2) -- (11.75,-3.2);
	\node[right of=wd] {\hspace{1.7cm}\scriptsize{dynamic world}};
	\draw[thick,fill=gray!50] (12,-4.90) circle (0.25cm) node (ws) {$w_s$};
	\draw[thick]  (10,-4.90) -- (11.75,-4.90);
	\node[right of=ws] {\hspace{1.3cm}\scriptsize{static world}};
	\draw[thick,<-] (10,-4.30) -- (10,-4.90);
	\draw[thick,->] (10,-3.2) -- (10,-3.8);
	\draw[thick,fill=gray!50] (10,-4.05) circle (0.25cm) node (w) {$w$};
	\node[right of=w] {\hspace{1.5em}\scriptsize{world}};
	\draw[thick,<-] (7.625,-4.05) -- (9.75,-4.05);
	
	\draw[white, text width=4cm] (12,-6.9) rectangle (14.2,-6.9) node[pos=0.5] {\includegraphics[width=2cm] {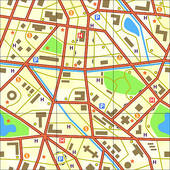}};
	\draw[white, text width=1.5cm] (13.8,-6.9) rectangle (14.2,-6.9) node[pos=0.5] {\scriptsize{\textit{\color{black}OpenStreetMap}}};
	\draw[double,<-] (12,-5.17) -- (12,-6.2);
	\draw[text width=3.7cm] (12.2,-5.3) rectangle (15.7,-6.1) node[pos=0.5] {\scriptsize{\texttt{\hspace{0.6em}cafe(name,GPS\_lat,GPS\_lon)\\\hspace{0.6em}atm(name,GPS\_lat,GPS\_lon)}}};
	
	\draw[thick] (3.75,-3.2) circle (0.25cm) node (params) {$\theta$};
	\node[above of=params] {\scriptsize{parameters}};
	\draw[thick,->] (3.75,-3.45) -- (3.75,-3.8);
	\draw[thick,fill=gray!50] (7.375,-4.05) circle (0.25cm) node (y) {$y$};
	\node (answer) [below of=y] {\scriptsize{retrieval}};
	\node (yDen) [below of=answer] {\scriptsize{\textit{\color{black}$y=\llbracket z \rrbracket_w$}}};
	\draw[thick] (3.75,-4.05) circle (0.25cm) node (z) {$z$};
	\node (logform) [below of=z] {\scriptsize{logical form}};
	\node (zProb) [below of=logform] {\scriptsize{\textit{\color{black}$z\sim p_\theta(z|x)$}}};
	\draw[thick,fill=gray!50] (0,-4.05) circle (0.25cm) node (x) {$x$};
	\node(xText) [below of=x] {\scriptsize{query}};
	\node(temp) [below of=xText] {\scriptsize{}};
	\node(eg) [below of=temp] {\scriptsize{Example:}};
	\draw[white, text width=3cm] (-0.5,-6) rectangle (2.5,-7) node[pos=0.5] {\scriptsize{\textit{\color{green!40!black}what is there on the right of the campus center?}}};
	\draw[white] (7.1,-6.5) rectangle (8,-7.5) node[pos=0.5] {\includegraphics[width=2.5cm] {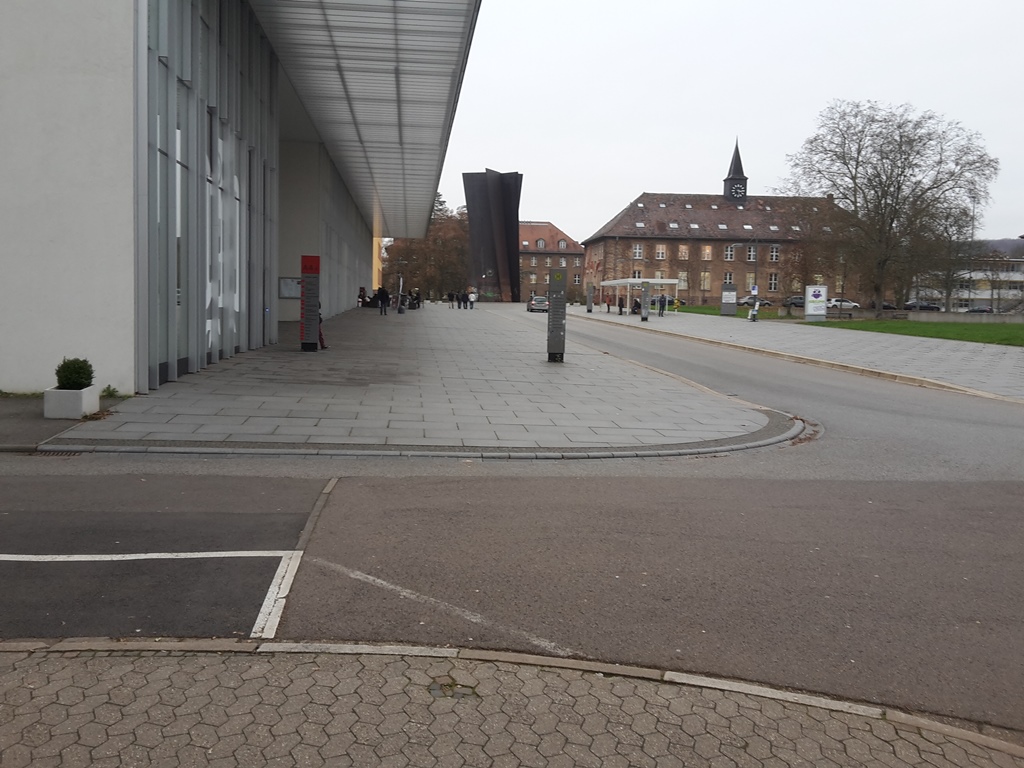}};
	\draw[thick,->] (0.25,-4.05) -- (3.5,-4.05) node[pos=0.5,above] {\textbf{\scriptsize{Semantic Parsing}}};
    \draw[thick,->] (4,-4.05) -- (7.125,-4.05) node[pos=0.5,above] {\textbf{\scriptsize{Interpretation}}};
    \draw (3.75,-6) node {*};
    \draw[] (3.75,-6.1) node {\tiny{\color{blue}1}};
    \draw (3.75,-6.2) -- (3.75,-6.4);
    \draw[] (3.75,-6.5) node (rel) {\tiny{\color{blue}1}};
    \draw[rounded corners, text centered] (3.25,-6.6) rectangle (4.25,-6.9) node[pos=0.5] {\tiny{\color{red}rightOf}};
    \draw[] (3.75,-7) node {\tiny{\color{blue}2}};
    \draw (3.75,-7.1) -- (3.75,-7.2);
    \draw[] (3.75,-7.3) node (rel) {\tiny{\color{blue}1}};
    \draw[rounded corners, text centered] (2.92,-7.4) rectangle (4.58,-7.7) node[pos=0.5] {\tiny{\color{red}campus\_center}};
\end{tikzpicture}
\caption{Our probabilistic graphical model: a question $x$ in natural language (a query from a user) is automatically mapped into a logical form $z$ by the semantic parser. It is next interpreted with respect to a world $w$ to give retrievals $y$. The world $w$ consists of a static part $w_s$ concretized as a database of geographical facts, and a dynamic part $w_d$ storing media content from \textit{Collective Visual Memory} and the user's spatio-temporal context (extracted from his/her metadata).}
\label{ProbModel}
\end{figure*}

\subsection{Learning-based, Contextual Media Retrieval by Semantic Parsing}

In this section, we describe how we approach learning-based, contextual media retrieval from natural language queries by a semantic parsing approach. First we describe the employed semantic parser architecture (inspired from ~\citet{Liang2012}) and show how to extend it towards a contextual media retrieval task. 
The probabilistic model of our architecture in shown in Figure~\ref{ProbModel}. A question $x$ (uttered by a user) is mapped to a latent logical form $z$, which is then evaluated with respect to a world $w$ (database of facts), producing an answer $y$. The world $w$ consists of $w_s$ (a static database of geographical information) and $w_d$ (a dynamic database which stores user metadata and information about media files in the \textit{Collective Visual Memory}). The logical forms $z$ are represented as labeled trees and are induced automatically from question-answer $(x, y)$ pairs.


\subsubsection{Question-Answering with Semantic Parsing}

We build our approach on a recently proposed framework for semantic parsing \citep{Liang2012} that has been shown to be able to answer questions about facts like geographical data and is trained solely on textual question-answer pairs. 
For example, the approach is capable of answering  question like \textit{``What are the major cities in California?''} with \textit{\{San Fransisco, Los Angeles, San Diego, San Jose\}} as an answer. In the semantic parsing framework (left part of Figure~\ref{ProbModel} labeled \textbf{\scriptsize{Semantic Parsing}} and \textbf{\scriptsize{Interpretation}}): `parsing' translates a question into its logical form $z$, and `interpretation' executes $z$ on the dataset of facts (word $w$) producing its denotation $\llbracket z \rrbracket_w$ - an answer.
Parameters $\theta$ are estimated solely on the training question-answer pairs $(x,y)$ with an EM algorithm maximizing the following posterior distribution:
\begin{equation}%
\label{eq:learning_semantic_parser}
\theta^{\ast} := \argmax_\theta \sum_{(x,y) \sim \mathcal{D}} \sum_z{\underbrace{\mathbf{1}\{y=\llbracket z \rrbracket_w\}}_{Interpretation}\hspace{0.2cm} \underbrace{p(z|x,\theta)}_{Semantic Parsing}}%
\end{equation}%
where $\mathcal{D}$ denotes a training set, $\mathbf{1}\{a=b\}$ is $1$ if a condition $a=b$ holds, and $0$ otherwise. The posterior distribution marginalizes over a latent set of valid logical forms $z$.
At test time, the answer is computed from the denotation $\llbracket z^{\ast} \rrbracket_w$ that maximizes the following posteriori:

\begin{equation}
\label{eq:test_time_semantic_parser}
  z^{\ast} := \argmax_{z} p(z|x,\theta^{\ast})
\end{equation}

The logical forms follow a dependency-based compositional semantics (DCS) formalism \cite{Liang2012} that consists of trees with nodes labeled with predicates and edges labeled with relations between the predicates. DCS is mainly introduced to efficiently encode feasible solutions. 

The underlying principle of the parsing is built on two components -- lexical semantics and compositional semantics. Lexical semantics learns a mapping from textual words into pre-defined predicates, and uses hand-designed lexical triggers that map specific parts-of-speech into a set of candidate predicates. Compositional semantics establishes relations between the predicates to generate the logical forms (DCS trees). The distribution over logical forms is modeled by a log linear distribution $p_\theta(z|x)\propto \mathrm{e}^{\phi(x,z)^T\theta}$, where the feature vector $\phi$ measure compatibility between the question $x$ and a logical form $z$. We perform a gradient descent scheme in order to optimize for parameters $\theta$. For a more detailed exposition of the semantic parser and parameter optimization in these models, we refer the reader to \citet{Liang2012}. In the following, we discuss our decomposition of the world $w$ into two parts: $w_s$ and $w_d$ (Figure~\ref{ProbModel}). 


\subsubsection{Static and Dynamic Worlds}
\label{dynamicExtension}

The existing works that use such a semantic parser are based on a static environment \citep{Liang2012, berant2014semantic, Malinowski2014}. In contrast to these, in our scenario a human user (the source of the query - $user$ in Figure~\ref{ProbModel}) relocates herself in space and time in a continuously changing environment. The pool of media content -- \textit{Collective Visual Memory} -- also grows as new media is added (by multiple users - crowd icon in Figure~\ref{ProbModel}). Such an environment leads us to our decomposition of the world $w$ into a static part $w_s$, which consists of geographical facts such as names of buildings or theirs GPS coordinates and inherits all the properties of the aforementioned previous works, and a dynamic and egocentric part $w_d$ (Figure~\ref{ProbModel}).

The dynamic world $w_d$ decomposes even further into $w_{d_m}$ that stores media metadata (timestamp, GPS coordinates) and is updated with continuously growing \textit{Collective Visual Memory}, and $w_{d_u}$ that models the user's context by storing her metadata (GPS coordinates, viewing direction). The latter is set anew for each query before it is fed into the semantic parser. Such representation renders the world $w = w_s + w_d$ static to the semantic parser although it is constantly changing.


\subsubsection{Modelling User's Context}
\label{context}

The user's context is modelled through predicates \texttt{person(LAT,LON,VIEW\_DIR)} 
where \texttt{LAT, LON, VIEW\_DIR} represent the user's current latitude, longitude and viewing directions respectively. These predicates are stored in the dynamic database of user metadata $w_{d_u}$ which is updated at query time for each query. 

Understanding egocentric spatial relations in natural language questions has for long intrigued the research community and forms a separate research area by itself~\cite{regier2001grounding, lan2012image, guadarrama2013grounding,892}. In our work, we approach ambiguity in the frame of reference \cite{Malinowski2014a} by defining predicates that resolve the spatial relations ``front of'', ``behind'', ``left of'' and ``right of'' based on the geomagnetic reference frame as well as the user-centric reference frame. Each of these spatial relations are modeled as two-argument predicates such as \texttt{frontOf(A,B), behind(A,B), leftOf(A,B)} and \texttt{rightOf(A,B)} where \texttt{A} denotes the GPS coordinates of the entity in question (extracted from $w_s$) and \texttt{B} denotes the GPS coordinates of the media files in $w_{d_m}$.

Similarly, the temporal references in questions (e.g. ``what happened here five days ago?'', ``how did this place look like in December?'') are modelled through the predicate \texttt{day(X)} where X is the referenced time-stamp (for example 20150511). These are resolved by mapping to trigger a predicate \texttt{view(A)}, where \texttt{A} is the list of media files having the same time-stamp as that in the predicate \texttt{day($\cdot$)}.

\begin{figure}
\centering
\includegraphics[width=0.9\columnwidth]{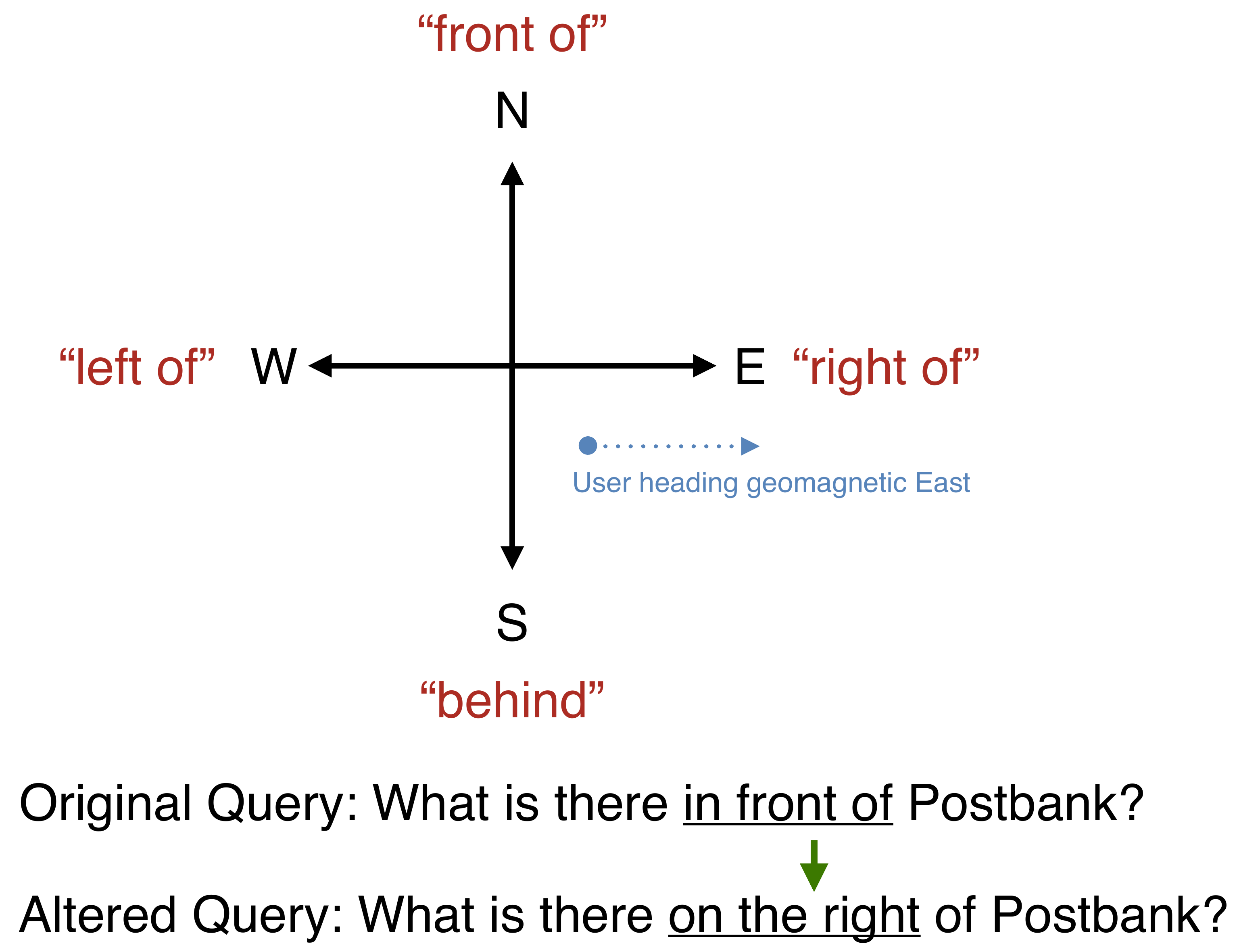}
\caption{Modification of spatial reference in query for integrating \textit{egocentrism} to media retrieval}
\label{refFrame}
\end{figure}

However, it is difficult to understand the hidden intent for contextual questions which includes an egocentric reference frame. This is because humans do not adhere to any consistent reference frame. They may consider their own physical ``left hand'' for ``left of'' or the physical ``left side'' of the geographical entity. The first possibility is tackled by programming the semantic parser to follow the geomagnetic reference frame. Then, with the assumption that the direction in which the human user faces is the local north, the spatial reference in the query is modified in a pre-processing step. This is explained in Figure~\ref{refFrame} -- if the user faces east and queries for ``What is there in front of postbank?'', the question would be changed during pre-processing to ``What is there on the right of postbank?''. The semantic parser would predict answer for this changed question. For simplicity we have narrowed down to only the four basic heading directions - north, south, east and west.

\subsubsection{Media Retrieval as Answers}
\label{section:media_retrieval_as_answers}

In contrast to previous work on question answering \cite{Liang2012,Malinowski2014}, we desire to retrieve media as answers to natural language questions instead of textual information. This can be modeled by generating references to media files as denotations of logical forms. For example, the question \textit{``What is there on the right of the campus center?''} would be transformed into the following symbolic representation (after training the system):\\
\texttt{answer(A, (rightOf(A,B), const(B, `campus\_center'))} with its
denotation \{`image12', `image58', `image234', ...\}, where  `image12', `image58' and so on are references to images with visual contents depicting geographical entities on the right of the university campus center. The answers are predicted with respect to a world which consists of 
the name, timestamp, GPS coordinates and the month of the media file acquisition. Once the denotations of a logical form are predicted, the actual media files are extracted from the \textit{Collective Visual Memory} (physically, a file system storing all captured media). These extracted media files are then returned to the user. Figure~\ref{qualResults} shows examples of retrieved results.


\subsection{Data Collection} %

To enable the spatio-temporal exploration of a certain geographic area we inherently require a database which record physical features on the ground along with their types (e.g. building, cafe, highway, etc.), names and GPS locations -- this constitutes our static world $w_s$. To support media retrieval we need a database of images and videos rich with metadata.
We also require natural language queries paired with corresponding media content as retrievals for training and testing our query-retrieval model. In the absence of a suitable benchmark, we needed to record our own data set, where the geographical facts ($w_s$) are obtained from OpenStreetMap, and the \textit{Collective Visual Memory} and query-retrieval pairs are collected from regular users.


\subsubsection{Geographical Facts}
\label{geoFacts}

OpenStreetMap~\cite{haklay2008openstreetmap} is a freely-available and well-documented collection of geographical data. The topological data structure used has four basic elements or \textit{data primitives} -- \textit{nodes}, \textit{ways}, \textit{relations} and \textit{tags}. Physical entities on the ground such as buildings, highways, ATMs, banks, restaurants etc. are registered in the map database in terms of these \textit{data primitives}. 
\begin{figure}
		\begin{subfigure}{\columnwidth}
		\centering
		\includegraphics[width=0.75\columnwidth]{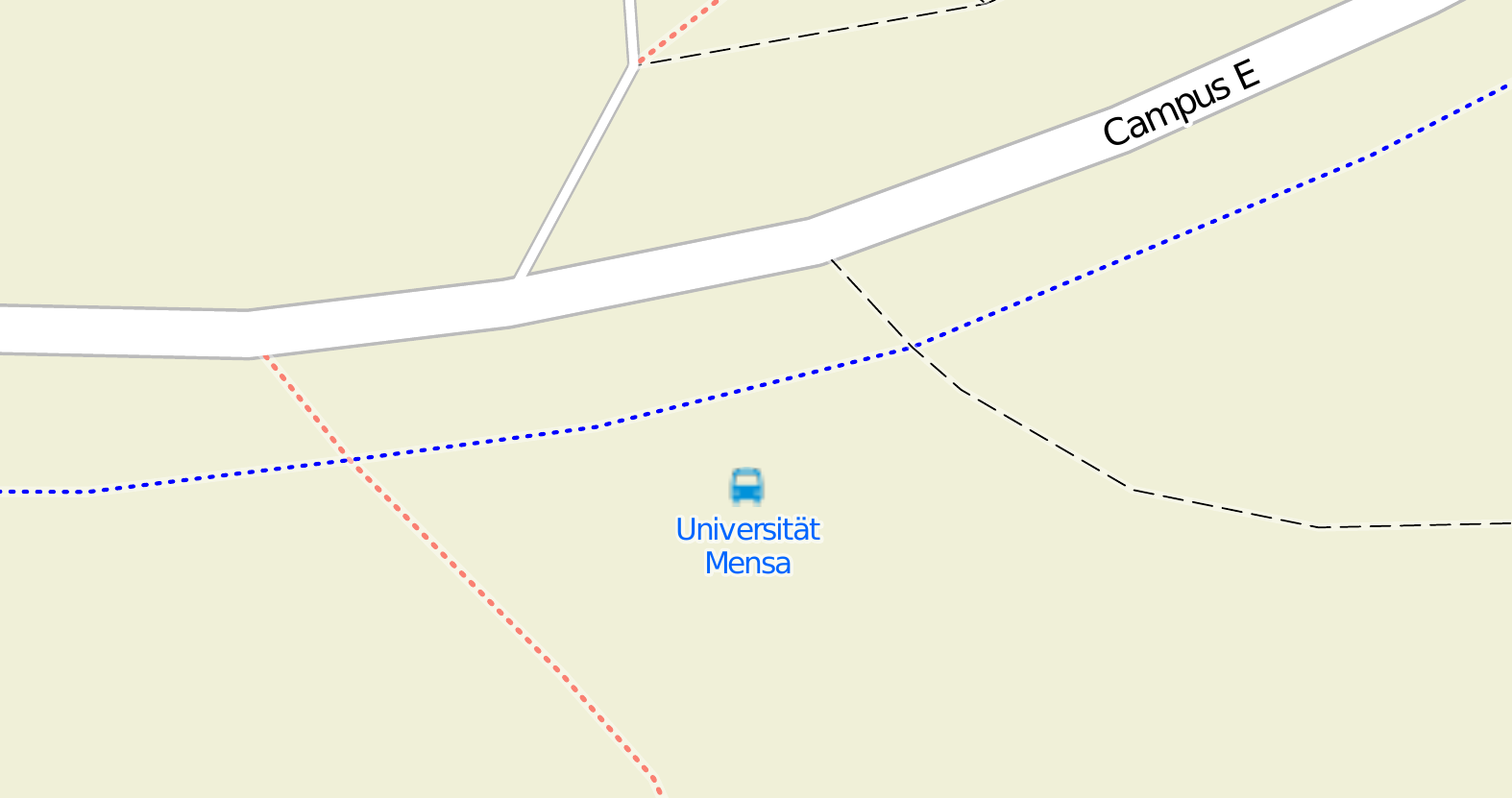}
		\caption{A section of the OpenStreetMap view of the university campus}
		\label{mapView}
		\end{subfigure}
\vspace{0.3cm}

		\begin{subfigure}{\columnwidth}
			\fbox{\begin{minipage}{0.95\columnwidth} 
			\small{$\langle$nodeid = ``344240596'' visible = ``true'' version = ``6'' changeset = ``9208001'' timestamp = ``2011-09-04T11 : 43 : 28Z'' user = ``arnhar'' uid = ``495739'' lat = ``81.24279'' lon = ``35.18783''$\rangle$\\
			$\langle$tagk = ``amenity'' v = ``bus\_stop''/$\rangle$\\
			$\langle$tagk = ``name'' v = ``Universit{\"a}t Mensa''/$\rangle$\\
			$\langle$/node$\rangle$
		
			}
			\end{minipage}}
			\caption{XML rendition of the physical entity shown in Figure~\ref{mapView}}
			\label{xml}
		\end{subfigure}
		\vspace{0.3cm}
		
		\begin{subfigure}{\columnwidth}
			\fbox{\begin{minipage}{0.95\columnwidth} 
			\small{
			bus\_stop(`universitaet\_mensa',49.2562752,7.0436771).
			}
			\end{minipage}}
			\caption{Entry in $w_s$ corresponding to the entity in Figure~\ref{xml}}
			\label{wsEntry}
		\end{subfigure}
		
\caption{Example of OpenStreetMap data}
\label{openStreetMap}
\end{figure}

In our study, we restrict the spatial scope of our system to a university campus. 
Figure~\ref{mapView} shows the map view of a part of the university campus depicting a physical entity on the ground -- a bus-stop named \textit{``Universit{\"a}t Mensa''}. The XML rendition of this part of the map (available for download) in shown in Figure~\ref{xml}. 
We used information such as the type of the physical entity (e.g. building, cafe, highway etc.), their names, and their GPS coordinates as our static database of facts ($w_s$) (Figure~\ref{wsEntry}).


\subsubsection{Collective Visual Memory}

Participants were asked to capture media (images and videos) at various locations of the university campus for a month using their mobile devices. In total our instance of the \textit{Collective Visual Memory} consists of 1025 images and 175 videos. Metadata such as GPS coordinates and time-stamp registered with each media file constitute our media database $w_{d_m}$.

The process of media acquisition was coupled with the collection of natural language questions. Participants were instructed to formulate a question and capture the photo(s)/ video(s) that they would expect as the corresponding answer. 1000 questions-answer pairs with spatial references were collected (one question could have multiple answers). Question-answer pairs with temporal references could not be collected because of the trivial infeasibility of capturing events from the past. The data set was randomized and divided into two parts -- 500 train questions and 500 test questions. To introduce sufficient amount of variations in natural language we chose participants from different cultural and linguistic background. 
We will make our data-set (the query-retrieval pairs, \textit{Collective Visual Memory} and the geographical facts) publicly available at time of publication.

\section{Experiments}

For our experiments we use the geographical facts and a \textit{Collective Visual Memory} as described in the previous section.We use a dataset consisting of query-retrieval pairs formulated by real-life users. It consists of user queries which follow no particular template and contains spatial relations in addition to those pre-defined as predicates, such ``near'', ``beside'', ``ahead of'', ``opposite to'' etc.

In this section we describe the experiments conducted, state their results and discuss our observations. We further propose the concept of personalization of a media retrieval system to adapt to specific user perceptions. Finally, we provide a qualitative assessment of the usefulness of our contextual media retrieval system.


\subsection{Evaluation of Learning Procedure}
\label{learning}

\begin{figure}
\centering
\includegraphics[width=0.9\columnwidth]{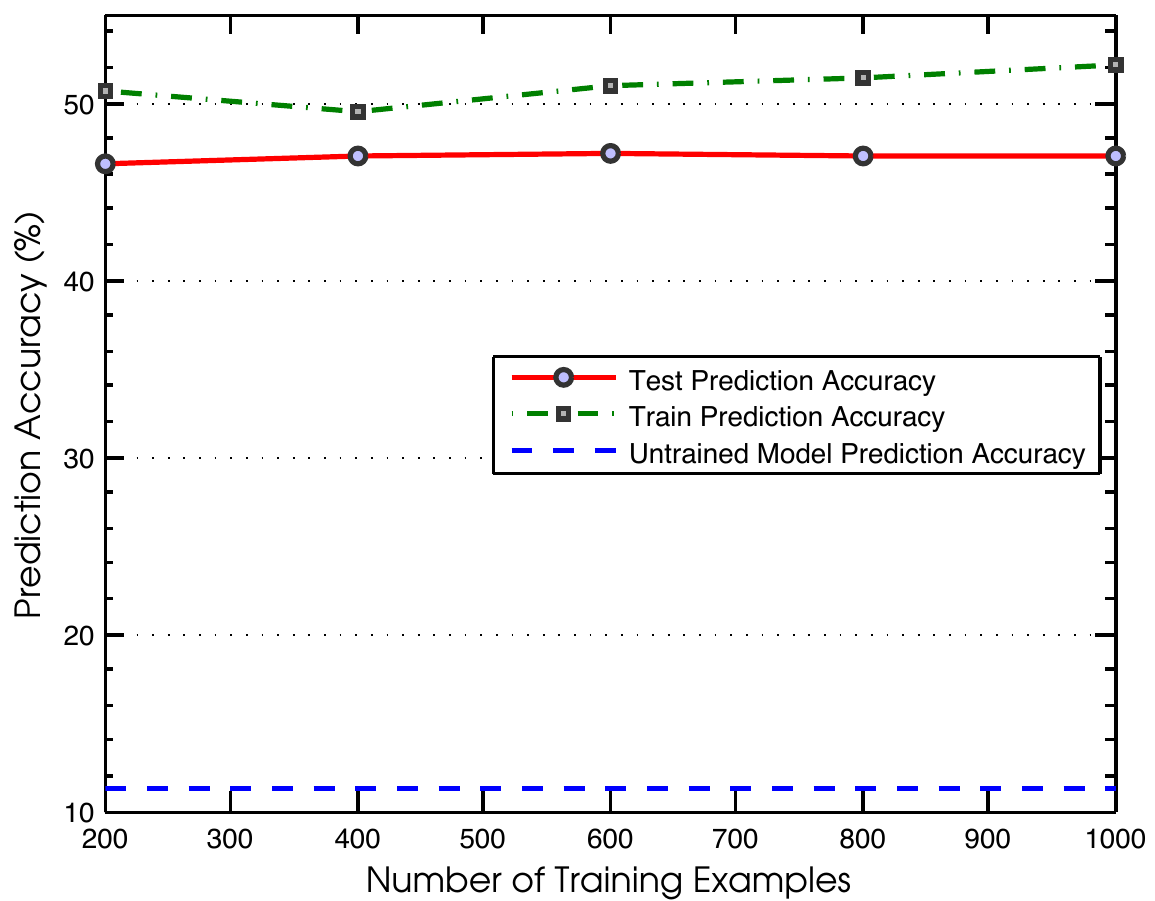}
\caption{Effect of increasing number of training examples on prediction accuracy}
\label{effectOfTraining}
\end{figure}

To study the effect of learning on prediction accuracy we first trained a model with synthetically generated query-retrieval pairs (SynthModel). The queries are generated by templates -- ``what is there \textit{$<$spatial relation$>$} of \textit{X}?'', ``what happened here \textit{Y} days/weeks/months/years ago?'', ``what did this place look like in \textit{Z}?'', where \textit{$<$spatial relation$>$} $\in$ \{``in front'', ``behind'', ``on the right'', ``on the left''\}, \textit{X} $\in$ \{names of buildings, cafes, restaurants etc.\}, \textit{Y} $\in$ \{natural numbers\} and \textit{Z} $\in$ \{names of months\}. The contextual cues \textit{`here', `this place'} are fixed to a particular location. The retrievals follow pre-defined rules to resolve spatial (according to the geomagnetic reference frame) and temporal relations. The untrained model was found to have a prediction accuracy of 11.23\%. 
We observe a strong improvement of performance to 46\% from as little as 200 training examples (Figure~\ref{effectOfTraining}).

Regular users use a variety of grammatical constructs as common in a spoken language. Therefore the queries collected from them were rich with a number of spatial relations not restrictive to the ones we represent as predicates (section~\ref{context}). Also, the answers to similar queries were subjective.
To account for the variability and subjectivity in this type of data, and to study the effect of learning on prediction accuracy, we trained a query-retrieval model on human queries and retrievals (HumanModel).
As before, we used a weakly supervised learning approach that only requires query-retrieval pairs without any supervision on the logical forms.
The model was trained through a human-in-the-loop training procedure using a relevance feedback mechanism. Since the human trainer was familiar with the geographic scope of our work, it was also possible to provide feedback on the retrievals of the temporal queries.
We found that during the training the query-retrieval model learned to associate different spatial relations to pre-defined predicates. For example, the parser has learned to map the spatial relation ``ahead of'' to the predicate \texttt{frontOf($\cdot$)}.

\begin{figure}%
	\centering
	\includegraphics[width=0.9\columnwidth]{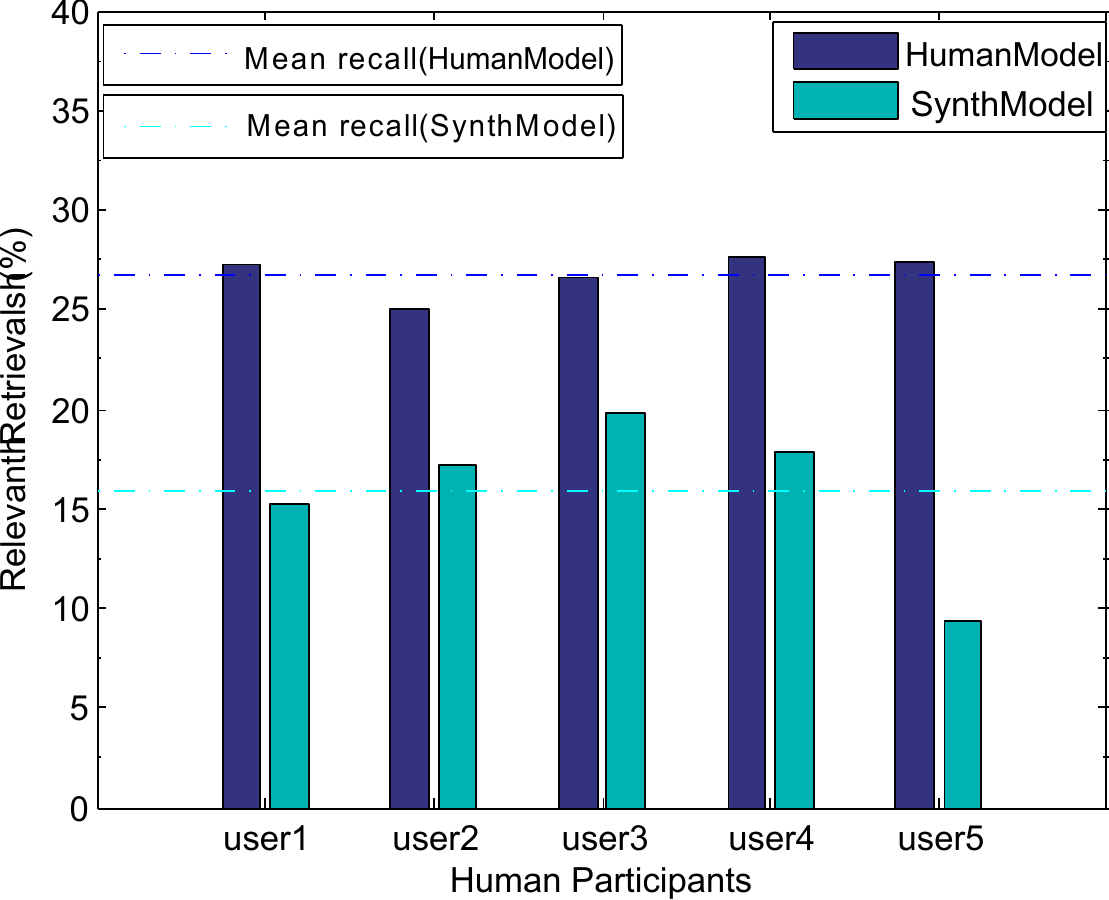}%
	\caption{Recall of HumanModel and SynthModel}%
	\label{recall}%
	\end{figure}

A comparison to the previous model shows that the HumanModel (26.67\%) yields greater recall than the SynthModel (15.88\%) on queries collected from humans (Figure~\ref{recall}), where recall is defined as the percentage of relevant retrievals among all test queries. This shows that our HumanModel is able to learn and adapt to the variations in natural language utterances and also interpret a variety of spatial relations in spoken queries. We use this model for our evaluations described in the following sections.

\subsection{Model Evaluation}
 
Since a contextual application calls for the involvement of prospective users and their satisfaction in using it, we decide on a qualitative assessment of the system.

Humans are inherently inconsistent in their perception of directions and idea of
reference frames~\cite{levinson2003space,Malinowski2014a}. The nature of understanding/speaking English questions also has variations based on a person's socio-cultural background. Hence, a system relying on fixed question templates and a particular set of rules to resolve spatial references does not guarantee high accuracy. A satisfactory result for one person may prove to be irrelevant for another. To better understand these perceptual biases and yet efficiently analyze the system, a series of user studies were conducted. 

\subsubsection{Evaluation of the Retrieved Results and Human Disagreements}

The goal of this user study is to observe how accurate regular users found our system. 
Five users were asked to evaluate the retrieved results for 500 test questions as ``relevant'' or ``irrelevant''. The study was conducted in a lab set-up. Users looked at retrieved results for each question on a computer screen and stated whether they find the retrievals relevant or irrelevant to the question. A canonical reference frame was used in this experiment to resolve spatial relationships in queries. According to this convention, ``front of'' meant ``north of'', ``behind'' meant ``south of'', ``right of'' meant ``east of'' and ``left of'' meant ``west of''. 

\begin{figure}%
\centering
\includegraphics[width=0.9\columnwidth]{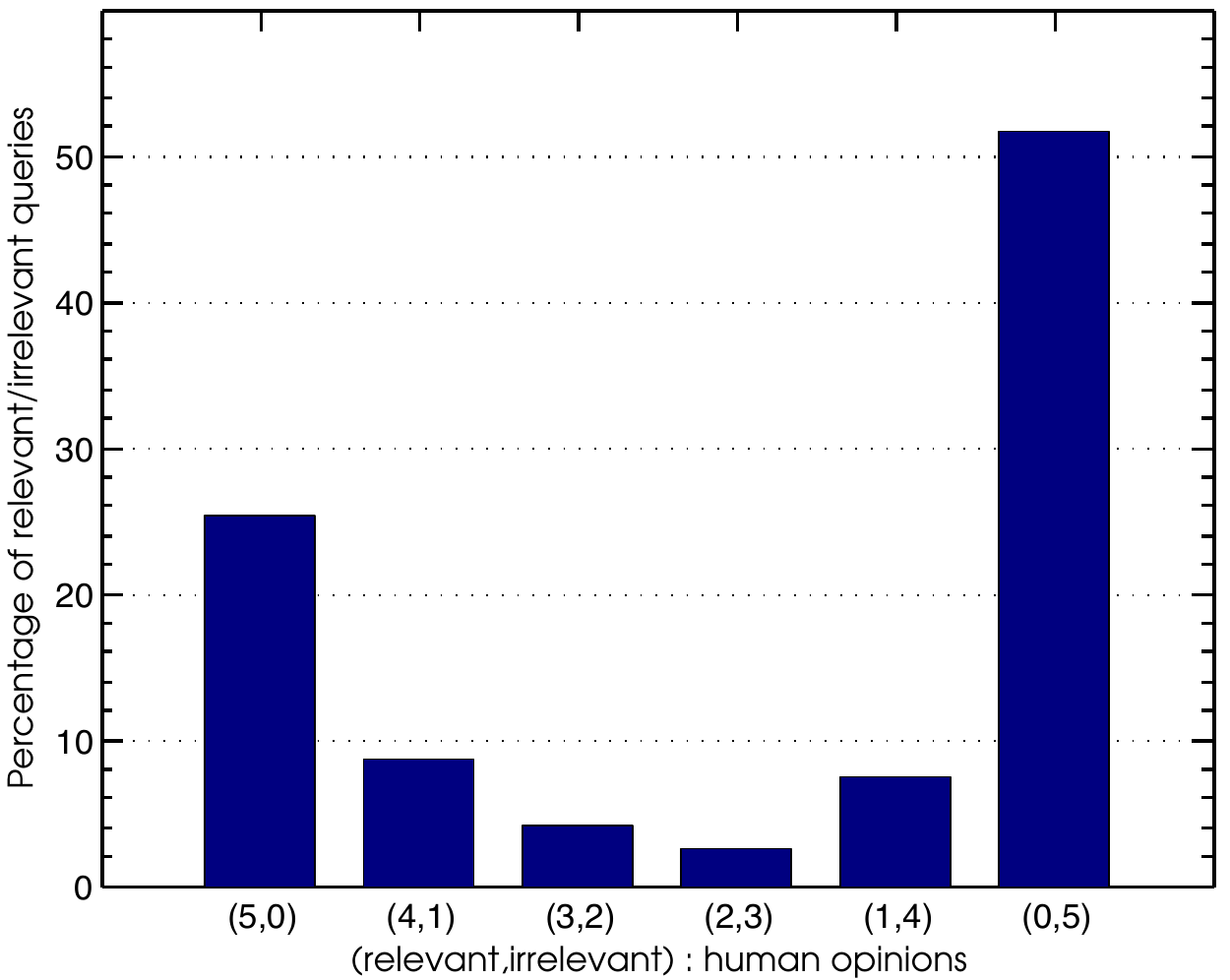}%
\caption{Inter-user variability in opinion}%
\label{agree-disagree}%
\end{figure}

We observed that for each question the opinions varied. Based on this observation we divide the test questions into six groups -- \textit{(5,0)}, queries for which all five users agreed that the retrievals were relevant; \textit{(4,1)}, queries for which four users found the retrievals relevant and one user found them irrelevant and likewise.
Figure~\ref{agree-disagree} depicts the result of this analysis. For 26.67\% of the queries all five users deemed the retrievals relevant. However, if we consider the cases in which most of the users found the retrievals relevant, this number rises to 40\%. 
The numbers in the middle region of the graph in Figure~\ref{agree-disagree} point out the prominent difference in opinions among participants. This accounts for about 25\% of all queries. We observed that the inter-user variability stems from the inherent inconsistencies with regards to reference frame resolution. This result also hints towards the difficulty of the problem at hand since satisfactory answers for one user may be unsatisfactory for others. The high agreement in the last column is because of some unavoidable factors - scanty media content (our geographic scope could not be well covered in images and videos due to lack of infrastructure), incorrect POS tagging (this resulted in incorrect retrievals type, for e.g. text), etc. 

From the observation from this user study -- that human disagreed in their opinion of relevance and irrelevance -- we conjecture that instead of using the geomagnetic reference frame, the use of user-centric reference frames for retrieving answers could improve the performance of the system. In the deployment of the user-centric reference frame we mean to follow the user's physical egocentric directions -- for example, her `right hand side' for ``right of'' etc. (explained in greater details in section~\ref{context}).

\subsubsection{Canonical and User-centric Reference Frame}

\label{canUserRefFrame}
In order to study the impact of using two different conventions of spatial relations resolution, we conducted this user study. Users were given two sets of retrieved results for each question -- one set of media files retrieved according to the geomagnetic reference frame and the second set retrieved according to the user-centric reference frame. The experimental settings are similar to the previous user study.

\begin{figure}%
\centering
\includegraphics[width=0.9\columnwidth]{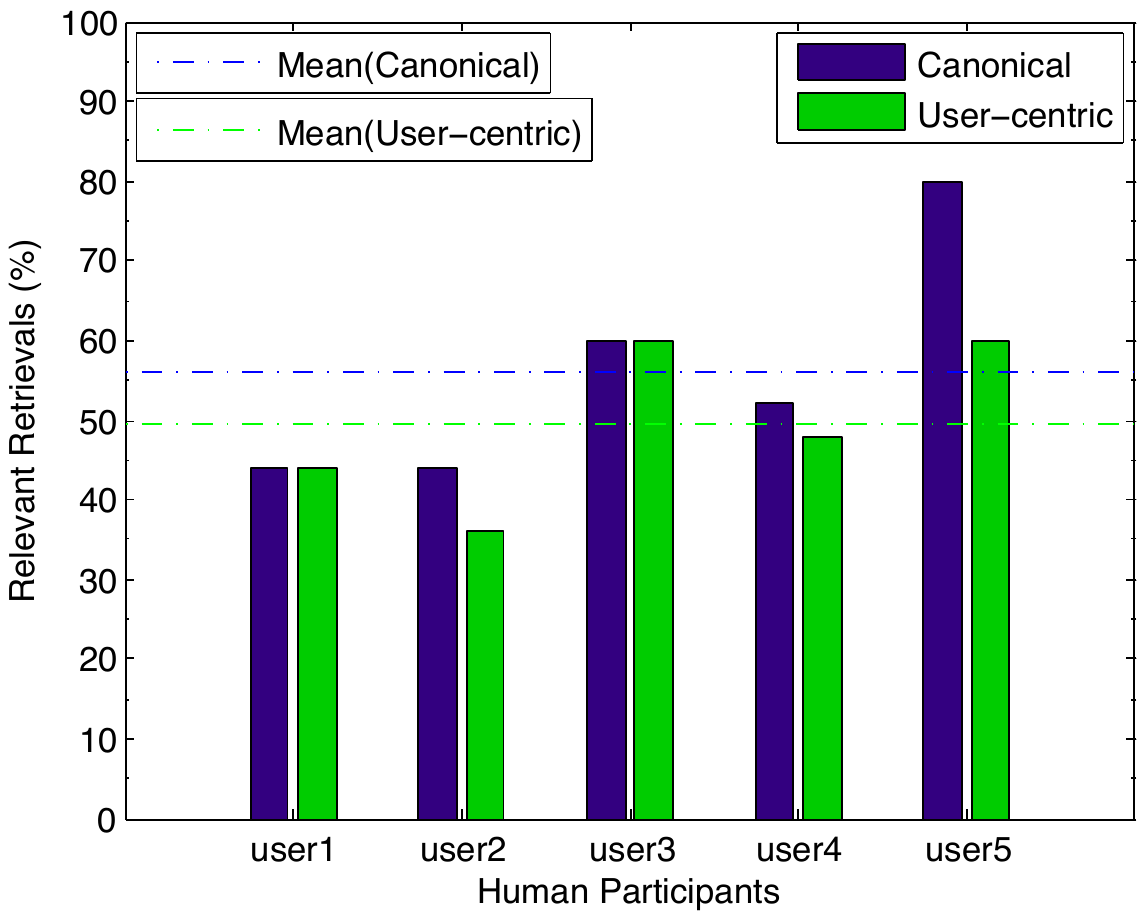}%
\caption{Difference in reference frame resolution among humans}%
\label{ref-frame}%
\end{figure}

Figure~\ref{ref-frame} shows the result of this user study. user1 and user3 remained neutral to the use of separate reference frames while the other users slightly preferred the canonical reference frame over the user-centric reference frame. This observation further highlights the subjectivity of the task.

\subsubsection{Personalization of Xplore-M-Ego} 

\begin{figure}
	\centering
	\begin{subfigure}{0.7\columnwidth}
			\bordermatrix{~ & U1 & U2 & U3 & U4 & U5 \cr
                  M1 & \colorbox{red!10}{27.7} & 9.6 & 16.23 & 13.35 & 17.27 \cr
                  M2 & 21.46 & \colorbox{red!10}{37.8} & 35.6 & 25.36 & 26.58 \cr
									M3 & 18.48 & 15.7 & \colorbox{red!10}{43.85} & 17.87 & 33.9 \cr
									M4 & 15.42 & 25.87 & 35.5 & \colorbox{red!10}{41.25} & 29.85 \cr
									M5 & 14.07 & 18.59 & 38.7 & 28.64 & \colorbox{red!10}{62.43} \cr}
			\caption{Precision}
			\label{personalization_precision}
	\end{subfigure}
	
	\begin{subfigure}{0.7\columnwidth}
			\bordermatrix{~ & U1 & U2 & U3 & U4 & U5 \cr
                  M1 & \colorbox{red!10}{23.39} & 8.2 & 13.68 & 11.25 & 14.56 \cr
                  M2 & 19.42 & \colorbox{red!10}{34.22} & 32.2 & 22.9 & 24.06 \cr
									M3 & 13.6 & 11.47 & \colorbox{red!10}{32.5} & 13.02 & 24.72 \cr
									M4 & 13.68 & 22.95 & 31.5 & \colorbox{red!10}{33.33} & 26.49 \cr
									M5 & 6.18 & 8.16 & 16.9 & 12.58 & \colorbox{red!10}{27.15} \cr}
		\caption{Recall}
		\label{personalization_recall}
	\end{subfigure}
	
	\begin{subfigure}{0.7\columnwidth}
		\bordermatrix{~ & U1 & U2 & U3 & U4 & U5 \cr
                  M1 & \colorbox{red!10}{25.36} & 8.84 & 14.84 & 12.21 & 15.79 \cr
                  M2 & 20.38 & \colorbox{red!10}{35.9} & 33.9 & 24.06 & 25.25 \cr
									M3 & 15.66 & 13.25 & \colorbox{red!10}{37.33} & 15.06 & 28.59 \cr
									M4 & 14.49 & 24.32 & 33.38 & \colorbox{red!10}{36.86} & 28.06 \cr
									M5 & 8.58 & 11.34 & 23.53 & 17.48 & \colorbox{red!10}{37.84} \cr}
		\caption{F\textsubscript{1}-score}
		\label{personalization_fscore}
	\end{subfigure}
\caption{Quantitative analysis of personalization of Xplore-M-Ego}
\label{personalization}
\end{figure}

Having observed this inter-person subjectivity, we hypothesize that personalization of our media retrieval system would increase its accuracy on a per user basis. The user study which we discuss in this section was conducted to investigate this hypothesis.

By using an online relevance feedback mechanism, five users ($U1,U2,U3,U4,U5$) were asked to train five different query-retrieval models ($M1,M2,$ $M3,M4,M5$) with 500 questions from the data-set collected from regular users. Every user was then asked to evaluate all five models keeping the identity of the model trained by each of them hidden.

The quantitative analysis of this study -- precision\footnote{$\text{precision} = \text{relevant}\ \text{retrievals/media retrievals}$}, recall\footnote{$\text{recall} = \text{relevant}\ \text{retrievals}/\text{total\ number\ of\ test\ queries}$} and F\textsubscript{1}-score\footnote{$F_1\ \text{score} = 2 * (\text{precision} * \text{recall})/(\text{precision} + \text{recall})$} -- are shown in Figure~\ref{personalization}. The diagonals show the user-specific evaluation results and the rows depict inter-user evaluation results. The difference in opinion among the users is very prominent, highlighting the challenge involved in the machine understanding of hidden human intent in natural language. Nonetheless, it is clear from the figure that users deemed their own models more accurate than those trained by others. This observation leads us to believe that the query-retrieval model can be trained over time through relevance feedback to adapt to user-specific preferences of spatial relation resolution -- hence, it should be personalized. This consolidates our hypothesis -- personalization of our media retrieval system increases its accuracy on a per user basis.

\subsubsection{User Experience Evaluation} 

To understand the usefulness of our contextual media retrieval system, we made an usability/desirability study. 10 participants were given the Google Glass installed with our client-side application and asked to walk around in the university campus while making voice queries that involve spatio-temporal references. Afterward they were asked to fill in the USE Questionnaire~\cite{lund2001measuring}. This questionnaire has four groups of questions -- Usefulness, Ease of Use, Ease of Learning and Satisfaction. Each question can be rated on a scale from 1 to 7, 1 meaning `strongly disagree' and 7 meaning `strongly agree'. 10 questions most representative of the entire questionnaire are chosen. The mean and standard deviation of the ratings of these questions are shown in Table~\ref{useQ}.
\begin{table}
\centering
\begin{tabular}{@{}p{6.57cm}p{0.5cm}p{0.5cm}@{}}
\toprule
USE Questionnaire                             &\hspace{-0.9em}Mean &SD\\\midrule
It is useful.                                    & 6.2       & 0.63        \\
It saves me time when I use it.                  & 6.1       & 0.73        \\
It is easy to use.                               & 6.3       & 0.48        \\
I can use it without written instructions.       & 5.8       & 1.22        \\
Both occasional and regular users would like it. & 5.4       & 1.42        \\
I learned to use it quickly.                     & 6.5       & 0.52        \\
I quickly became skillful with it.               & 6.1       & 0.99        \\
I am satisfied with it.                          & 5.5       & 0.52        \\
It is fun to use.                                & 6.3       & 0.82        \\
I would recommend it to a friend.                & 5.9       & 0.73        \\
\bottomrule
\end{tabular}
\caption{User Experience Evaluation: Mean Rating and Standard Deviation. The grades are between 1 ('strongly disagree') and 7 ('strongly agree')}\label{useQ}
\end{table}
The result of this evaluation shows that regular users strongly agree that our contextual media retrieval application is useful in daily life. Moreover, they find the application easy to use, very easy to learn and they are satisfied with the outcome.

\section{Discussion}

Due to the complex nature of our problem that involves natural language queries, media and map data, human concepts, in particular of spatial and temporal language, and complex contextual cues, we have faced a wide range of challenges. We highlight three of these in this section and discuss limitations and future work.\\

\noindent
{\bf Frame of reference:}
Proper spatial resolution is required in a successful communication with machines. Unfortunately, there is no a unique frame of reference, and hence even a simple statement involving ``left of'' has different meanings for different users. However, our findings suggests two promising research directions in the reference frame resolution task. First, our inspection of the user study shows that the users often resolve spatial relations in spoken language for the navigation task according to the frontal direction of the physical object (from observations of section \ref{canUserRefFrame}). Hence, a suitable map database that stores information about the frontal direction of the objects would help. We are unaware of such database or efforts to augment existing map data with such meta information. Second, our study on personalized Xplore-M-Ego suggests a more individual approach where the architecture learns to understand spatial relations by interacting with the user. While our online learning approach shows a first promising step in this directions, more complex models of person specific biases and shared notions across users could further improve the learning.\\

\noindent
{\bf Diversity of Named Entities:}
Our approach uses a static database that contains information about the geographical entities, for instance the name of the entity, extracted from the OpenStreetMap. However, the participants in our study use a number of different names to refer to the same entity -- the formal full name, an acronym, a popular name, or even a name in a different language. Handling such diversity is a complicated task for the semantic parser, and hence we resort to manually adding all possible common names for each entity in the database. However, such human annotation may still be incomplete. An alternate method for handling the coverage of the database is to use suitable knowledge-bases containing acronyms and regional names of geographical entities or crawling additional web resources. To the best of our knowledge, such information about synonyms of map entities is currently not pursued, but would greatly benefit applications that relate to map data such as ours.\\

\noindent
{\bf Scalability: }
The program induction step of the semantic parser, where a logical form $z$ is searched over a large space of possible predicates and theirs relations (Eq. \ref{eq:learning_semantic_parser} and \ref{eq:test_time_semantic_parser}), is computationally demanding, and does not scale well with a large number of predicates representing geographical facts.
We deal with this problem by reducing the spatial scope to a university campus. In deployment, we envision a system that directly works in a spatial scope of the user, and updates the database by geographical facts in $w_s$ while the user is relocating in space and time.

\section{Conclusion} 

In this paper we proposed Xplore-M-Ego -- a novel system for media retrieval using spatio-temporal
natural language queries in a dynamic setting. Our work brings forth a new direction to this paradigm by exploiting a user's current context. Our approach is based on a semantic parser that infers interpretations of the natural language queries. We contribute several extensions which enable the user to dynamically refer to his/her context by spatial and temporal concepts. We further analyzed the system in the various user studies that highlight the importance of our adaptive and personalized training approaches.

\bibliographystyle{abbrvnat}
\bibliography{contextualMediaRetrieval}  

\end{document}